\documentclass[useAMS,usenatbib]{mn2e}

\usepackage[dvips]{graphicx}

\title[Magnetorotational supernovae  with jets]
{Magnetorotational supernovae with jets }
\author[S.G.Moiseenko, G.S.Bisnovatyi-Kogan, and N.V.Ardeljan]
{S.G.Moiseenko$^{1}$\thanks{E-mail: moiseenko@iki.rssi.ru (SGM)},
G.S.Bisnovatyi-Kogan$^{1}$\thanks{gkogan@iki.rssi.ru (GSBK)} and
N.V.Ardeljan$^{2}$\thanks{ardel@cs.msu.su (NVA)}\\
$^{1}$Space Research Institute, Profsoyuznaya str. 84/32, Moscow
117997, Russia \\
 $^{2}$Department of Computational Mathematics and
Cybernetics, Moscow State University, Vorobjevy Gory, Moscow
B-234, Russia }

\begin{document}

\date{Accepted  . Received ; in original form }

\pagerange{\pageref{firstpage}--\pageref{lastpage}} \pubyear{2005}

\maketitle

\label{firstpage}

\begin{abstract}
We present results of 2D simulation of magnetorotational (MR)
supernova accompanied by jet formation in the core collapse
supernova explosion. Initial magnetic field used in the
simulations has dipole-like symmetry. Contrary to the simulations
of MR supernova with initial quadrupole-like magnetic field, where
the matter was ejected mainly near the equatorial plane, in
presence of the dipole-like initial magnetic field  the supernova
explosion is developing preferably along a rotational axis, and
leads to formation of a protojet. We expect that protojet
propagation through the envelope of the star  will be accompanied
by its collimation. The magnetorotational instability (MRI) was
found in simulations, similar to the earlier considered case of
the quadrupole-like initial magnetic field. Our estimations show
that the characteristic time for the reconnection of the magnetic
field is much larger than the MRI development time. The supernova
explosion energy for the dipole-like field is about $0.61\cdot
10^{51}$erg, and about $\>0.13M_\odot$ of mass was ejected during
the explosion.

\end{abstract}


\section{Introduction}
The idea of the MR mechanism of the core collapse supernova,
suggested by \citet{bk1970},  was to get energy for the supernova
explosion from the  rotation of the star due to magnetic field
action. During the collapse stellar rotation is becoming
differential i.e. central parts of the star rotate much faster
than the envelope. 1D numerical simulations made by
\citet{{bkpopsam}}, \citet{abkp}
had shown that the differential rotation in
presence of magnetic field leads to appearing of the toroidal
component of the magnetic field,
 its linear growth,
 and following
formation of MHD shock wave. In 1D approach the
rotation was differential along $r$ axis only (the star in 1D was
a cylinder) what means that the only $H_r$ component of the
magnetic field was winded up and produced the toroidal
($H_\varphi$) component. In a 2D case the
system has more degrees of freedom and the rotation is
differential not only along the $r$ axis. The differential
rotation is using both components of the magnetic field ($H_r, \>
H_z$) for the toroidal field amplification. 2D simulations of MR
mechanism with the initial magnetic field of the quadrupole-like
type of symmetry have shown (\citet{abkm2000}, \citet{mbka2004},
\citet{abkm2005}) that the amplification of the toroidal component
of the magnetic field leads to  formation of a compression wave,
which moves along steeply decreasing density profile and quickly
transforms into the fast MHD shock wave. Due to the
quadrupole-like initial magnetic field the supernova explosion was
developed predominantly along the equatorial plane. The explosion
energy for that case was about $0.6 \cdot 10^{51}$erg. The value
of the ejected energy is enough for the explanation of the core
collapse supernova explosion. The main qualitative difference
between 1D (\cite{abkp}) and 2D (\cite{abkm2005}) MR supernova
simulation is an appearance of the magnetorotational instability
(MRI) which significantly reduces the  time of the growth of the
toroidal magnetic field up to MR supernova explosion. In 1D case
the MRI is not developed due to the small number of degrees of
freedom.

The first   simulations of the MR processes in stars have been
done by \citealt{leblanck}, after which MR processes in the stars
in relation to the core collapse supernova explosion  had been
simulated by \citealt{bkpopsam}, \citealt{abkp},
\citealt{mueller}, \citealt{ohnishi}, \citealt{symbalisty}.
Recently the interest to the MR processes (especially in
application to the core collapse supernova) was recommenced due to
increasing number of observational data about asymmetry of the
explosion, and possible collimated ejecta in connection with
cosmic gamma ray bursts (\citealt{abkm2000}; \citealt{akiyama};
\citealt{yamada}; \citealt{takiwaki}; \citealt{kotake2004};
\citealt{abkm2005}; \citealt{yamasaki}). In the paper by
\cite{bktut} it was estimated that a significant part of the type
Ib,c supernovae could explode as magnetorotational supernovae.

In the present paper we describe the results of the 2D simulations
of the MR supernova explosion with the initial dipole-like
magnetic field. The qualitative difference between our simulation
of the same problem with the initial quadrupole-like magnetic
field is the formation of mildly collimated protojet, directed
along the rotational axis. As in the paper by \cite{abkm2005} we
have found a development of the  MRI leading to rapid explosion
even at a relatively weak initial magnetic field. In the
simulations we have used an ideal MHD with the infinite
conductivity. In spite of the fact that we supposed the
conductivity to be infinite, in reality it is very big, but
finite. That means that during winding up the force lines of the
magnetic field and the MRI development, the reconnection of the
magnetic field  lines is possible. If the reconnection time would
be smaller or comparable to the characteristic time of the MRI
development then the reconnection process could reduce
effectiveness of the MR supernova mechanism.

We have done estimations for the reconnection time and have found
that it is much larger  than the time of the development of the
MRI, and will not influence significantly on the MR supernova
explosion (MRE). MRI in application to the core collapse supernova
was studied  by \cite{spruit}, \cite{akiyama}.

For the simulations we have used implicit Lagrangian numerical
method on triangular grid of variable structure. This method was
used for the simulations of a number of different astrophysical
problems:  simulations of the collapse of cold rapidly rotating
protostellar clouds (\cite{abkkm1996}), magnetorotational
processes for the collapsing magnetized protostellar cloud
(\cite{abkm2000}), core collapse and formation of the neutron star
(\cite{abkkm2004}), MR supernova simulations with quadrupole-like
initial magnetic field (\cite{abkm2005}).

\section{Formulation of the problem}

\subsection{Equations of state and neutrino losses}\label{sect_neut}

For the simulations we use the equation of state from
\citet{abkpch}:

\begin{eqnarray*}
  P \equiv P(\rho,T)=P_0(\rho)+\rho \Re T + \frac {\sigma T^4} {3},
\end{eqnarray*}

\begin{equation}\label{pressure}
P_0(\rho)=\left\{
\begin{array}{l}
P_0^{(1)}=b_1\rho^{5/3}/(1+c_1\rho^{1/3}),0 \leq\rho\leq\rho_1,\\
P_0^{(k)}=a\cdot 10^{b_k({\textrm{lg}}\rho-8.419)^{c_k}},\rho_{k-1}\leq\rho\leq\rho_k,\\
\hskip 4cm k=\overline{2,6}\\
\end{array}
\right.
\end{equation}
\[
\begin{array}{lll}
b_1=10^{12.40483}   & c_1=10^{-2.257} & \rho_1=10^{9.419}    \\
b_2=1.           & c_2=1.1598      & \rho_2=10^{11.5519}  \\
b_3=2.5032       & c_3=0.356293    & \rho_3=10^{12.26939} \\
b_4=0.70401515   & c_4=2.117802    & \rho_4=10^{14.302}   \\
b_5=0.16445926   & c_5=1.237985    & \rho_5=10^{15.0388}  \\
b_6=0.86746415   & c_6=1.237985    & \rho_6\gg \rho_5     \\
a=10^{26.1673}.  &                 &
\end{array}
\]
Here $\Re$ is the gas constant taken equal to $0.83\times 10^8
{cm^2}{s^{-2}}K$, $\sigma$ is the constant of the radiation
density, $P$ is pressure, $\rho$ is density, and $T$ is
temperature. In the expression $P_0(\rho)$ the value $\rho$ was
identified with the total mass-energy density. For the cold
degenerated matter the expression for $P_0(\rho)$ is the
approximation of the tables from \citet{bps, mjb}.

In the neighborhood of points $\rho=\rho_k$ in equation
(\ref{pressure}) the function $P_0(\rho)$ was smoothed in the same
way as in \citet{abkpch}, to make continuous the derivative
 ${\textrm d}P_0/{\textrm d} \rho$ :

\begin{equation}\label{smoothp}
P_0(\rho)=\left\{
\begin{array}{l}
P_0^{(k)}, \rho\in[\rho_{k-1}+\xi_{k-1},\rho_k-\xi_k],\\
\hskip 1cm k=\overline{1,6}, \> \rho_0+\xi_0=0,\\
\theta_k P_0^{(k)}+(1-\theta_k P_0^{k+1}),\\
\hskip 1cm \rho\in[\rho_k-\xi_k,\rho_k+\xi_k],  k=\overline{1,5},\\
\end{array}
\right.
\end{equation}
where
$$
\theta_k=\theta(\rho)=\frac{1}{2}-\frac{1}{2}\sin\left(\frac{\pi}{2\xi_k}(\rho-\rho_k)\right),
\quad \xi_k=0.01\rho_k.
$$

The specific energy (per mass unit) was defined thermodynamically
as:
\begin{equation}\label{inten}
\varepsilon=\varepsilon_0(\rho)+\frac{3}{2} \Re T +\frac{\sigma
T^4}{\rho}+\varepsilon_{Fe}(\rho,T).
\end{equation}

The value $\varepsilon_0(\rho)$ is defined by the relation
\begin{equation}\label{inten0}
\varepsilon_0(\rho)=\int\limits_0^\rho
\frac{P_0(\tilde{\rho})}{\tilde{\rho}^2}\textrm{d}\tilde{\rho}.
\end{equation}

 The term  $\varepsilon_{Fe}$ in equation (\ref{inten}) is responsible for iron dissociation.
 It is used in the following form:
\begin{equation}\label{ferrum0}
  \varepsilon_{Fe}(\rho,T)=\frac{E_{b,Fe}}{A\>m_p}
  \left(\frac{T-T_{0Fe}}{T_{1Fe}-T_{0Fe}}\right).
\end{equation}
It is supposed that in the region of the iron dissociation the
iron amount is about $50$ per cent of the mass, $E_{b,Fe}=8\times
10^{-5}$erg is the iron binding energy, $A=56$ is the iron atomic
weight, and $m_p=1.67\times 10^{-24}$g is the proton mass,
$T_{0Fe}=0.9\times 10^{10}$K, $T_{1Fe}=1.1\times 10^{10}$K. For
the numerical calculations formula (\ref{ferrum0}) has been
slightly modified (smoothed):
\begin{equation}\label{ferrum1}
  \varepsilon_{Fe}(\rho,T)=\frac{E_{b,Fe}}{A\>m_p}\frac{1}{2}
 \left\{1+\sin\left[\pi  \left(\frac{T-T_{0Fe}}{T_{1Fe}-T_{0Fe}}\right)-\frac{\pi}{2}\right]\right\}.
\end{equation}

The neutrino losses for Urca processes are used in the form, taken
from \citet{bkpopsam}, approximating the table of \citet{iin}:
\begin{equation}\label{urca}
  f(\rho,T)=\frac {1.3 \cdot 10^9 {\textrm {\ae}}(\overline{T})\overline{T}^6}
  {1+(7.1\cdot 10^{-5}\rho \overline{T}^3)^{\frac{2}{5}}}\quad
  {\textrm {erg}} \cdot {\textrm{g}}^{-1} \cdot {\textrm {s}}^{-1},
\end{equation}

\begin{equation}
{\textrm {\ae(T)}}=\left\{
\begin{array}{rclcccccc}
1,&\overline{T}<7,\\
664.31+51.024 (\overline{T}-20), & 7\leq \overline{T} \leq 20,\\
664.31, & \overline{T}>20,
\end{array}
\right.
\end{equation}

$$\overline {T}=T\cdot 10^{-9}.$$

The neutrino losses from pair annihilation, photo production, and
plasma were also taken into account. These types of the neutrino
losses have been approximated by the interpolation formulae from
\citet{schindler}:
\begin{equation}\label{dopneu}
  Q_{tot}=Q_{pair}+Q_{photo}+Q_{plasm}.
\end{equation}
The three terms in equation (\ref{dopneu}) can be written in the
following general form:
\begin{equation}\label{schindler1}
  Q_d=K(\rho,\alpha)e^{-c\xi}\frac{a_0+a_1\xi+a_2\xi^2}{\xi^3+b_1\alpha+b_2\alpha^2+b_3\alpha^3}.
\end{equation}
For $d=pair$, $K(\rho,\alpha)=g(\alpha)e^{-2\alpha}$,
$$
g(\alpha)=1-\frac{13.04}{\alpha^2}+\frac{133.5}{\alpha^4}+\frac{1534}{\alpha^6}+\frac{918.6}{\alpha^8};
$$
For $d=photo$, $K(\rho,\alpha)=(\rho/\mu_Z)\alpha^{-5};$\\
For $d=plasm$, $K(\rho,\alpha)=(\rho/\mu_Z)^3;$
$$
\xi=\left(\frac{\rho / \mu_Z}{10^9}\right)^{1/3} \alpha.
$$

Here, $\mu_Z=2$ is the number of nucleons per electron.
Coefficients $c,\> a_i,$ and $b_i$ for the different losses are
given in the Table \ref{tabb1} of \citet{schindler}.
\begin{table*}
 \centering
 \begin{minipage}{140mm}
  \caption{Coefficients for equation (\ref{schindler1}) from \citet{schindler}.}\label{tabb1}
   \begin{tabular}{|l|l|l|l|l|l|l|l|}
  \hline
       & $a_0$ & $a_1$ & $a_2$ & $b_1$ & $b_2$ & $b_3$ & $c$ \\ \hline
  \multicolumn{8}{|c|}{$10^8\>K \leq T \leq \> 10^{10}\> K$}\\ \hline
  pair  & 5.026(19) & 1.745(20) & 1.568(21) & 9.383(-1) & -4.141(-1) & 5.829(-2) & 5.5924 \\
  photo & 3.897(10) & 5.906(10) & 4.693(10) & 6.290(-3) & 7.483(-3) & 3.061(-4) & 1.5654 \\
  plasm & 2.146(-7) & 7.814(-8) & 1.653(-8) & 2.581(-2) & 1.734(-2) & 6.990(-4) & 0.56457 \\ \hline
  \multicolumn{8}{|c|}{$10^{10}\>K \leq T \leq \> 10^{11}\> K$}\\ \hline
  pair  & 5.026(19) & 1.745(20) & 1.568(21) & 1.2383 & -8.1141(-1) & 0.0 & 4.9924 \\
  photo & 3.897(10) & 5.906(10) & 4.693(10) & 6.290(-3) & 7.483(-3) & 3.061(-4) & 1.5654 \\
  plasm & 2.146(-7) & 7.814(-8) & 1.653(-8) & 2.581(-2) & 1.734(-2) & 6.990(-4) & 0.56457 \\ \hline
\end{tabular}
\end{minipage}
\end{table*}
The general formula for the neutrino losses in a nontransparent
star has been written in the form, used by \citet{abkkm2004}:
\begin{equation}\label{neuttot}
  F(\rho,T)=(f(\rho,T)+Q_{tot})e^{-\frac{\tau_\nu}{10}}.
\end{equation}
The multiplier $e^{-\frac{\tau_\nu}{10}}$ in equation
(\ref{neuttot}), where $\tau_\nu=S_\nu n l_\nu$, restricts the
neutrino flux for non zero depth to neutrino interaction with
matter $\tau_\nu$. The cross-section for this interaction $S_\nu$
was presented in the form:
$$S_\nu=\frac{10^{-44}T^2}{(0.5965 \cdot 10^{10})^2},$$
and the concentration of nucleons  is
$$n=\frac{\rho}{m_p}.$$
The characteristic length-scale $l_\nu$, which defines the depth
for the neutrino absorbtion, was taken to be equal to the
characteristic length of the density variation as:
\begin{equation}\label{depth}
l_\nu=\frac{\rho}{ |\nabla \rho|}=\frac{ \rho}{\left((\partial
\rho /\partial r)^2+ (\partial \rho / \partial z)^2\right)^{1/2}}.
\end{equation}
The value $l_\nu$ monotonically decreases when moving to the
outward boundary; its maximum is in the centre. It approximately
determines the depth of the neutrino-absorbing matter. The
multiplier $1/10$ in the expression $e^{-\tau_\nu\frac{1}{10}}$
was applied because in the degenerate matter of the hot neutron
star only some of the nucleons with the energy near Fermi
boundary, approximately $1/10$, take part in the neutrino
processes.

\subsection{Basic equations} Consider a set of
magnetohydrodynamical equations with self-\-gra\-vi\-ta\-tion and
infinite conductivity:
\begin{eqnarray}
\frac{{\rm d} {\bf x}} {{\rm d} t} = {\bf v}, \nonumber \\
\frac{{\rm d} \rho} {{\rm d} t} +
\rho \nabla \cdot {\bf v} = 0,  \nonumber\\
\rho \frac{{\rm d} {\bf v}}{{\rm d} t} =-{\rm grad}
\left(P+\frac{{\bf H} \cdot {\bf H}}{8\pi}\right) + \frac {\nabla
\cdot({\bf H} \otimes {\bf H})}{4\pi} -
\rho  \nabla \Phi, \nonumber\\
\rho \frac{{\rm d}}{{\rm d} t} \left(\frac{{\bf H}}{\rho}\right)
={\bf H} \cdot \nabla {\bf v},\> \Delta \Phi=4 \pi G \rho,
\label{magmain}\\
\rho \frac{{\rm d} \varepsilon}{{\rm d} t} +P \nabla \cdot {\bf
v}+\rho F(\rho,T)=0,
 \nonumber\\
P=P(\rho,T),\> \varepsilon=\varepsilon(\rho,T). \nonumber
\end{eqnarray}
here $\frac {\rm d} {{\rm d} t} = \frac {\partial} {
\partial t} + {\bf v} \cdot \nabla$ is the total time
derivative, ${\bf x} = (r,\varphi , z)$, ${\bf
v}=(v_r,v_\varphi,v_z)$ is the velocity vector, $\rho$ is the
density, $P$ is the pressure,  ${\bf H}=(H_r,\> H_\varphi,\> H_z)$
is the magnetic field vector, $\Phi$ is the gravitational
potential, $\varepsilon$ is the internal energy, $G$ is
gravitational constant, ${\bf H} \otimes {\bf H}$ is the tensor of
rank 2, and $F(\rho,T)$ is the rate of neutrino losses.

$r$, $\varphi$, and  $z$ are spatial Lagrangian coordinates, i.e.
$r=r(r_0,\varphi_0,$ and $z_0,t)$,
$\varphi=\varphi(r_0,\varphi_0,z_0,t)$, and
$z=z(r_0,\varphi_0,z_0,t)$, where $r_0,\varphi_0,z_0$ are the
initial coordinates of material points of the matter.

Taking into account symmetry assumptions ($ \frac \partial
{\partial \varphi} = 0$), the divergency of the tensor ${\bf H}
\otimes {\bf H}$ can be presented in the following form:
$$
{\rm \nabla\cdot}({\bf H} \otimes {\bf H})= \left(\begin{array}{l}
\frac {1}{r} \frac {\partial(rH_rH_r)}{\partial r} + \frac
{\partial(H_zH_r)} {\partial z}-
\frac {1}{r} H_\varphi H_\varphi \\
\frac {1}{r} \frac {\partial(rH_rH_\varphi)}{\partial r} + \frac
{\partial(H_zH_\varphi)} {\partial z}+
\frac {1}{r} H_\varphi H_r \\
\frac {1}{r} \frac {\partial(rH_rH_z)}{\partial r} + \frac
{\partial(H_zH_z)} {\partial z}
\end{array}
\right).
$$

Axial symmetry ($\frac \partial {\partial \varphi}=0$) and
symmetry to the equatorial plane  are assumed. The problem is
solved in the restricted domain. At $t=0$ the domain is restricted
by the rotational axis $r\geq 0$, equatorial plane $z\geq 0$, and
the outer boundary of the star where
 the density of the matter is zero, while poloidal
components of the magnetic field $H_r$, and $H_z$ can be non-zero.

 At
the rotational axis ($r=0$) the following boundary conditions are
defined: $(\nabla \Phi)_r=0,\> v_r=0$. At the equatorial plane
($z=0$) the boundary conditions are: $(\nabla \Phi)_z=0,\> v_z=0$.
At the outer boundary (boundary with vacuum) the following
condition  is defined:
 $P_{\textrm {outer boundary}}=0$.

We avoid explicit calculations of the function
$\varepsilon_0(\rho)$ in equation (\ref{inten0}), because this
term is eliminated from (\ref{magmain}) due to adiabatic equality:
\begin{equation}\label{adiab}
\rho\frac{{\rm d} \varepsilon_0}{{\rm d} t}=
-\frac{P_0}{\rho}\frac{{\rm d}\rho}{{\rm d} t}= P_0 \nabla \cdot
{\bf v},
\end{equation}
determining the fully degenerate part of the equation of state.
Therefore, we define
\begin{eqnarray*}
  \varepsilon^*=\frac{3}{2}\Re T +\frac{\sigma T^4}{\rho}+\varepsilon_{Fe}(\rho,T),\\
  P^*=\rho \Re T +\frac{\sigma T^4}{3}.
\end{eqnarray*}
The equation for the internal energy in equation (\ref{magmain})
can be written in the following form:

\begin{equation}\label{eqmod}
\rho\frac{{\rm d} \varepsilon^*}{{\rm d} t}+P^* \nabla\cdot{\bf
v}+\rho F(\rho,T)=0.
\end{equation}

\subsection{Initial magnetic field}
\begin{figure}
\centerline{\includegraphics{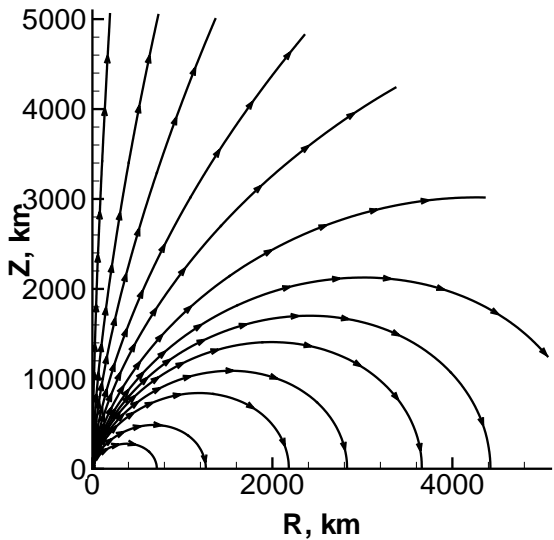}}
 \caption{The initial poloidal dipole-like magnetic field configuration.}
  \label{inimag}
\end{figure}

The initial poloidal magnetic field is defined similar to our
previous papers (\citet{abkm2005}, \citet{abkm2000}) by the
toroidal current $j_\varphi$ using Bio-Savara law. The  toroidal
current which determines the initial magnetic field should be
defined in the upper and in the lower hemispheres. The  field with
the quadrupole-like symmetry is formed by the toroidal current
antisymmetrical to the equatorial plane. The dipole-like magnetic
field is formed by  current symmetrical to the equatorial plane.
To define the initial poloidal magnetic field we have used the
following toroidal current $j_\varphi$ (in nondimensional
variables:
\begin{equation}
j_\varphi=\left\{
\begin{array}{l}
j_\varphi^u \> {\rm for} \> z>,< 0;\> \left(r-0.15\right)^2+0.2(z)^2\leq 0.3^2, \\
0 \> \> \> {\rm for} \> {\phantom {z\leq 0,}}
\>\left(r-0.15\right)^2+0.2(z)^2> 0.3^2,
\end{array} \right.
\label{inicur}\\
\end{equation}
where
\begin{eqnarray*}
j_\varphi^u  = A_j\times {\rm cos}\left[{\frac{\pi}{2}}
\left({\frac{(r-0.15)^2+0.2z^2}{0.3^2}} \right)\right]\nonumber\\
j_\varphi^d = j_\varphi^u.\nonumber
\end{eqnarray*}
The coefficient $A_j$
serves for adjusting the initial magnetic energy.
In the dipole-like type of symmetry (\ref{inicur}) there is
$H_r=0$ at the equatorial plane ($z=0$) (Fig. \ref{inimag}).
\begin{figure}
\centerline{\includegraphics{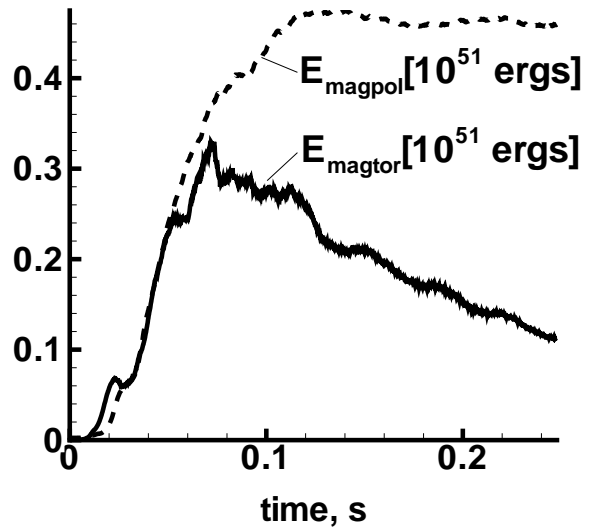}}
 \caption{Time evolution of the toroidal (solid line) and poloidal
 (dashed line) parts of magnetic
 energy
 during MR explosion with dipole-like initial magnetic field.}
  \label{magen}
\end{figure}
It is important to have a  force-free or a balanced initial
magnetic field. Using of the nonbalanced initial magnetic field
leads to appearance of artificial flows due to  "turning on"
effect. We start the simulation of the MagnetoRotational Explosion
(MRE) from the "equilibrium" state to avoid numerical problems,
arising from the nonstationarity, especially in outer region of
the core. These regions give a small input into the total energy
balance of the explosion.

To construct a configuration, where magnetic forces are balanced
with other forces (i.e. gravitational force, gradient of gas
pressure and centrifugal force) we  follow the procedure from
\citet{abkm2000} and \citet{abkm2005}.  The ratio between the
initial magnetic  and  gravitational energies was chosen to be
equal to $10^{-6}$, like in \citet{abkm2005}, for better
comparison of the present results with the results of simulations
of MR explosion with the initial quadrupole-like magnetic field.
The initial poloidal magnetic field in the center of the star at
start of the evolution of the toroidal magnetic field is $\sim
3.2\times 10^{13}$G.

\subsection{Core collapse simulation}\label{coll}

We first have done simulation of the rotating nonmagnetized core
collapse (\citealt{abkkm2004}). Initially the ratios between the
rotational and gravitational energies and between the internal and
gravitational energies of the star had been chosen as:
$$
\frac{E_{rot}}{E_{grav}}=0.0057,\quad
\frac{E_{int}}{E_{grav}}=0.727.
$$

The rotating core collapse resulted in formation of the proto
neutron star with an extended envelope. The bounce shock which was
formed during the collapse, goes through the envelope. At the end
of calculations in \cite{abkkm2004} at $t=0.2565$ s the bounce
shock reaches the outer boundary of our computational domain. The
shock leads to the ejection of $0.041$ per cent of the core mass
and 0.0012 per cent ($2.960\times 10^{48}$ erg) of the
gravitational energy of the star. The amounts of the ejected mass
and energy are too small to explain the core collapse supernova
explosion. After the core collapse ($t=0.261$ s) we obtain a
differentially rotating configuration.

As in previous simulation of the MRE  we divide the whole process
into three separate parts. The first part is the simulation of the
core collapse. The second part is the construction of the balanced
star with the  dipole-like magnetic field. The third part is the
calculation of MRE of this configuration.

The initial model for simulation of MRE was taken from the paper
of \cite{abkkm2004}, where the first part of the process
(collapse) was calculated. It is represented by differentially
rotating hot proto neutron with central density $\sim 2\times
10^{14}\> g/cm^3$, central temperature $3.5\times 10^{10}\> K$,
and angular velocity near the center, corresponding to the period
$P\sim 10^{-3}s$. At equatorial radius $\sim 10 \> km$, containing
$\sim 18\%$ of the mass the rotational period was equal to
$1.5\times 10^{-3}\>s$, and increased up to $\sim 35 \>s$ at the
outer boundary at the equator. The collapse calculations in the
paper of \cite{abkkm2004} have been started from rigidly rotating
iron  "white dwarf" star with the mass $\sim 1.2M_\odot$, 20\%
larger than Chandrasekhar limit for the pure iron star, with the
rotational period $P\approx 0.4 \> s$. The outer boundary is
expanding during the collapse, leading to large increase of the
rotational period.

After the core collapse and formation of a steady state
differentially rotating configuration, the initial poloidal
magnetic field, defined by the toroidal current (\ref{inicur}) was
'turned on'. As in the paper by \citet{abkm2005}, at the second
step the toroidal component was switched off, to obtain a balanced
magnetized configuration, and the evolution of only poloidal
components ($H_r,\> H_z$) of the magnetic field was permitted.

\section{Numerical method}
The numerical method used in present simulations  is based on the
implicit operator-difference completely conservative scheme on the
Lagrangian  triangular grid of variable structure. The
implicitness of the applied numerical scheme allows to make large
time steps. It is important to use implicit scheme in such kind of
problems due to the presence of two strongly different timescales.
The small timescale is defined by the huge sound velocity in the
central parts of the star. The big time scale is defined by the
characteristic timescale of the evolution of the magnetic field.
Conservation properties of the numerical scheme are important for
the exact fulfillment of the energy balance and divergence-free
property of the magnetic field.

During the simulation of the MRE the time step for the implicit
scheme was $\sim 10 \div 300$ times larger than time step for the
explicit scheme (CFL stability condition). It means that the total
number of time steps is $10\div 300$ times less than it would have
been done for explicit scheme,  what allows us to decrease time
approximation error. We did not make direct comparison of CPU time
per time step for our implicit scheme and an explicit scheme. Our
estimations show that  the total number of arithmetic operations
for the implicit scheme is $\approx 20$  times larger than it is
required for explicit scheme. Considerable decrease of the
required number of time steps leads to corresponding reduce of the
time approximation error.

 Grid reconstruction procedure
applied here for the reconstruction of the triangular lagrangian
grid is used both for the correction of the "quality" of the grid
and for the dynamical adaptation of the grid.

The method applied here  was developed, and its stability was
investigated in the papers by \citet{arko}, \citet{arkoche} and
references therein. It  was tested thoroughly with different tests
(\citet{abkm2000}).

\section{Results}

\subsection{Magnetorotational supernova explosion and protojet
formation}

After the core collapse we get a steady differentially rotating
configuration which properties are described in the section
\ref{coll}. The ratio of the rotational energy and gravitational
energy is equal to $0.073$. The star remains in such a state
pretty long time. We have done about 10000 time steps (it
corresponds to 0.03 s of physical time) and did not notice any
significant changes of the parameters of the star. While the
inclusion of even weak initial poloidal magnetic field leads to
 drastic changes.

After formation of the balanced configuration with the poloidal
dipole-like magnetic field, at the moment of 'switching on' the
equation for the evolution of the toroidal magnetic field we start
counting the time anew.

The toroidal magnetic field component
appearing due to the differential rotation is amplifying  with
time. The fastest amplification of the $H_\varphi$ takes place in
the regions of the maximum of the value $rH\cdot{\rm
grad}(V_\varphi/r)$ in the equation of the evolution of
$H_\varphi$. At the initial stage
$H_\varphi$
grows linearly with time (correspondingly toroidal magnetic energy
$H_\varphi^2/(8\pi)$ grows as a quadratic function)
(Fig.\ref{magen}). Due to
 appearance of the MRI (which will be described in the following
section),
after the stage of
a linear growth,
the toroidal magnetic field
 starts to grow much faster
(exponentially) . At this stage the poloidal magnetic field also
begins
the exponential growth.

The toroidal magnetic energy reaches its
maximal value $\approx 3.2 \times 10^{50}$ erg at $t=0.07$ s. This
value is $\approx 66\%$ of the maximal value of the  toroidal
energy in the case of initial quadrupole-like magnetic field,
where this maximum was reached at $t=0.12$ s (\citet{abkm2005}).
The maximal values of the toroidal magnetic field $H_\varphi
\approx 1.8\cdot 10^{16}$ G (in the case of the quadrupole this
maximal value was $\sim 2.5 \times 10^{16}$ G) are reached at the
distance
$\sim 10$ km from the center of the star.

After
reaching its maximal value the toroidal magnetic energy decreases
with the time.  At the developed stage the poloidal magnetic
energy
reaches $\approx 4.5\times 10^{50}$ erg,
and does not show any substantial decrease with time.

At $t=0.07$s a compression wave appears moving from the central
parts of the star due to
 increasing of the magnetic pressure.
 Moving along
 a steep density profile the compression wave
transforms
 into the fast MHD shock wave. The development of the
MHD shock outwards leads to the ejection of
mass and energy.

During the MR explosion a part  of the rotational energy ($\approx
10\%$) is transforming into the radial kinetic energy (which could
be related to supernova explosion energy). The neutrino transport
effects are taken into account approximately in flux limited
approximation variant (see section \ref{sect_neut}). A larger
fraction of the rotational energy is lost by neutrino emission,
while most part of the angular momentum is carried away by the
ejected matter.

\begin{figure}
  \includegraphics{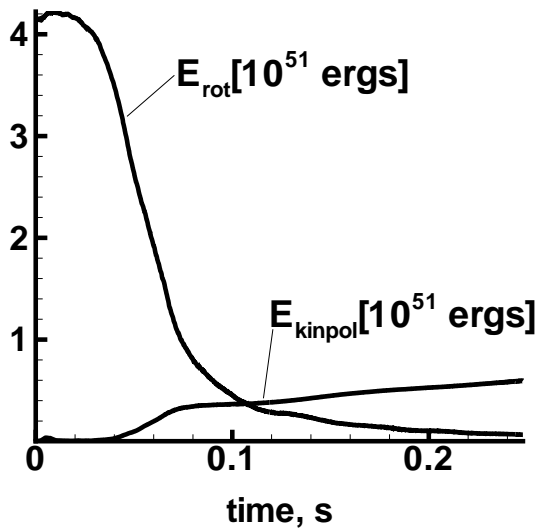}\\
  \caption{Time evolution of the rotational and radial kinetic energy
  during MR explosion.}
  \label{rotkin}
\end{figure}

\begin{figure}
  \includegraphics{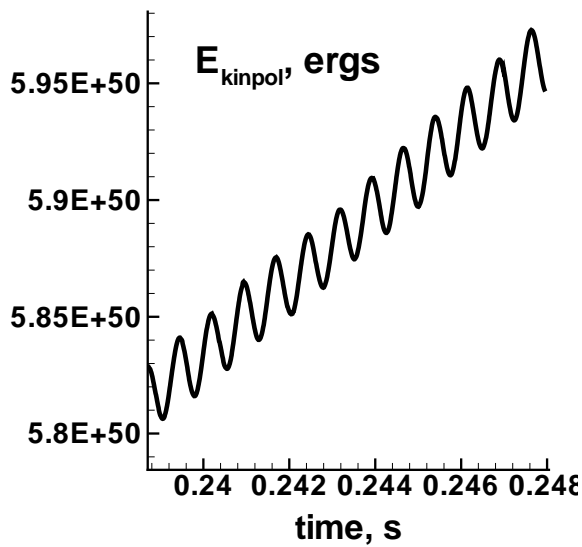}\\
  \caption{Zoomed part of the time volution of the radial kinetic energy
  (Fig. \ref{rotkin})
  during MR explosion.}
  \label{osc}
\end{figure}

At the Fig.\ref{rotkin} the time evolution of the rotational and
radial (poloidal) energies is represented.
At the final stage of the MRE simulations, the excitation of
eigenmodes of the proto neutron star happens.  The plot of the
radial kinetic energy (Fig. \ref{osc}) shows that this eigenmode
has a period of  $\approx 1$ ms.

\begin{figure}
  \includegraphics{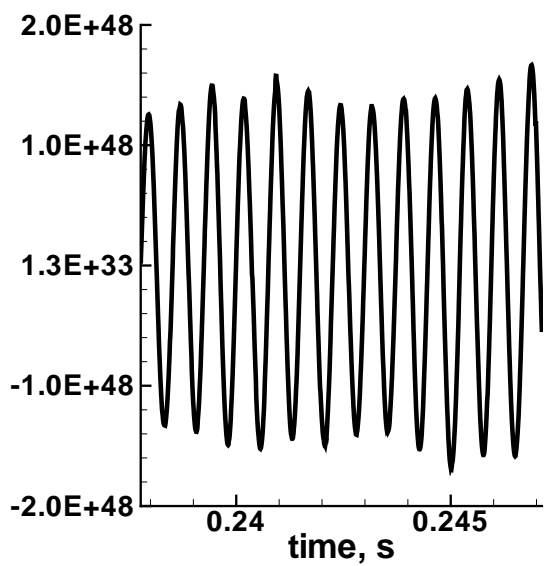}\\
  \caption{The time volution of the kinetic energy of
  the core only during MR explosion.}
  \label{coreosc}
\end{figure}

At the Fig. \ref{coreosc} the kinetic energy of the core is
represented. This plot was made by subtracting the kinetic energy of
the matter expanding by the supernova shock from the total kinetic
energy of the star.
The amplitude of the kinetic energy of the core oscillations is
$\approx 1.3 \cdot 10^{48}$ ergs. In the paper by \citet{burrows} a
mechanism of core-collapse supernova explosions was suggested,
based on the acoustic energy extracted from the core.

Our simulations have shown that the amount of the acoustic energy
generated in the collapsing core is much less than the expected
energy of core collapse supernova, note that our mathematical
method was different, and we have used more simplified description
of the physical processes than \citet{burrows}

\begin{figure}
  \includegraphics{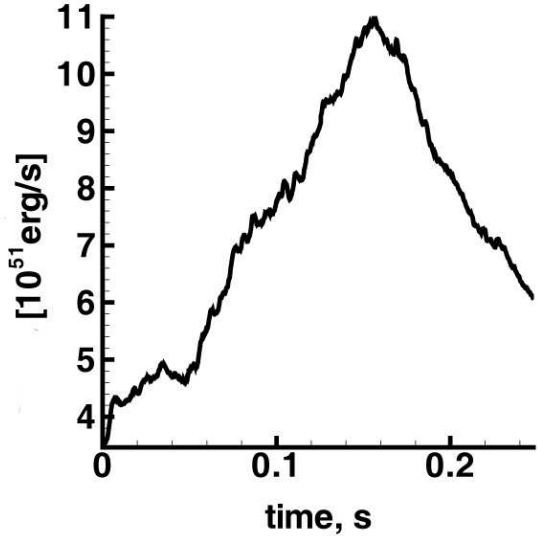}\\
  \caption{Time dependence of the neutrino luminosity $\int^{M_{core}}_0 F(\rho,T){\rm d}m$ during
  MR explosion.}\label{neutlum}
\end{figure}
\begin{figure}
  \includegraphics{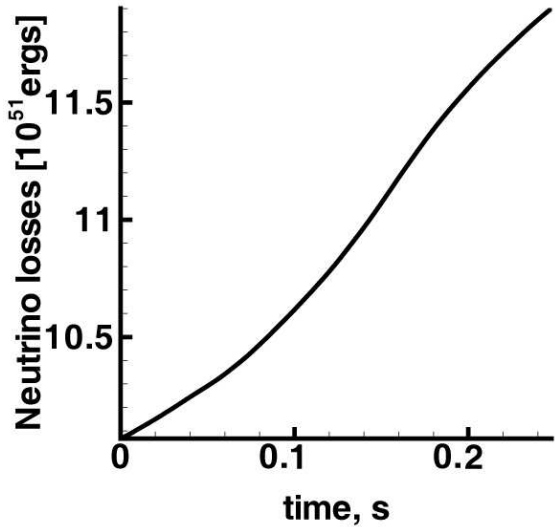}\\
  \caption{Time dependence of the neutrino losses during
  MR explosion.}\label{neutlos}
\end{figure}
At the Figs.\ref{neutlum}, \ref{neutlos} the time evolution of the neutrino
luminosity and neutrino losses during
MR explosion are represented.

\begin{figure}
\centerline{\includegraphics{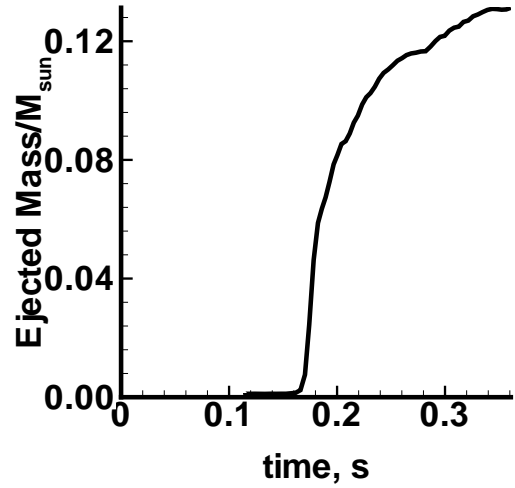}}
 \caption{Time evolution of the ejected mass during MR explosion.}
  \label{maseject}
\end{figure}
\begin{figure}
\centerline{\includegraphics{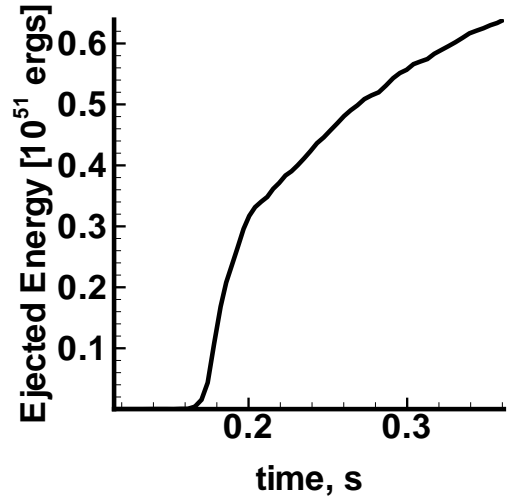}}
 \caption{Time evolution of the ejected energy during MR explosion.}
  \label{eneject}
\end{figure}
The amounts of ejected mass ($\approx 0.13M_\odot$) and energy ($6.1
\times 10^{50} $erg) are close to the corresponding values in the
case of MR supernova explosion with the initial quadrupole-like
magnetic field.  The
plots of the time evolution of the ejected mass and energy for the
dipole field are represented at Fig.\ref{maseject} and
Fig.\ref{eneject}.

\begin{figure}
  \centerline{\includegraphics{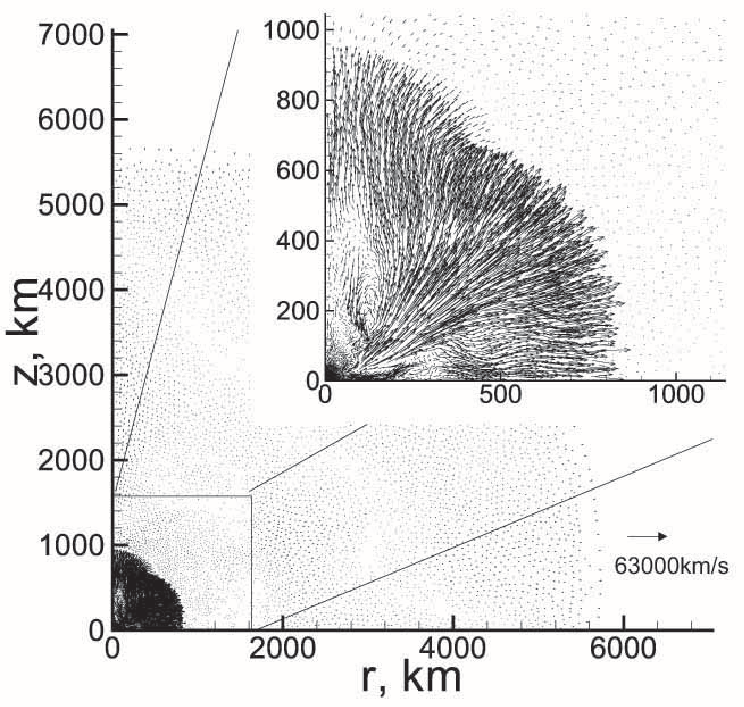}}
  \vspace{-0.2cm}
  \centerline{\includegraphics{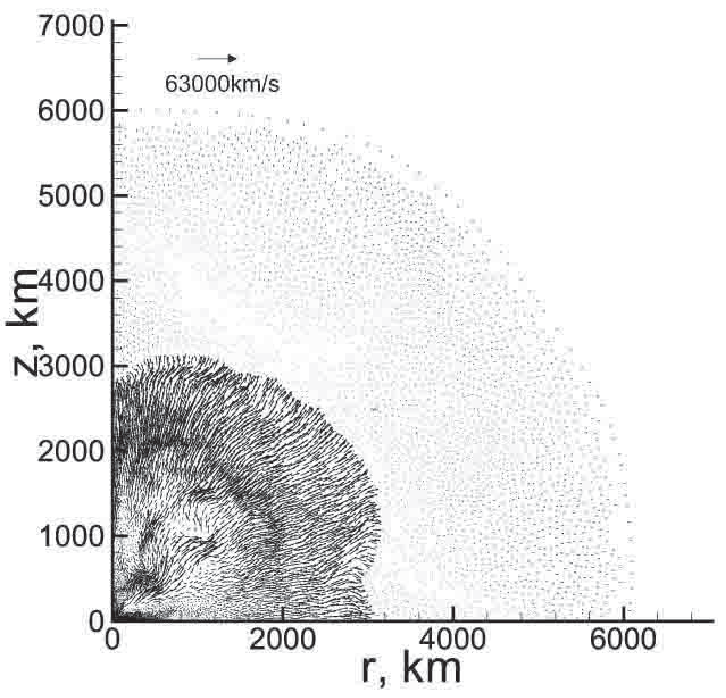}}
  \vspace{-0.2cm}
  \centerline{\includegraphics{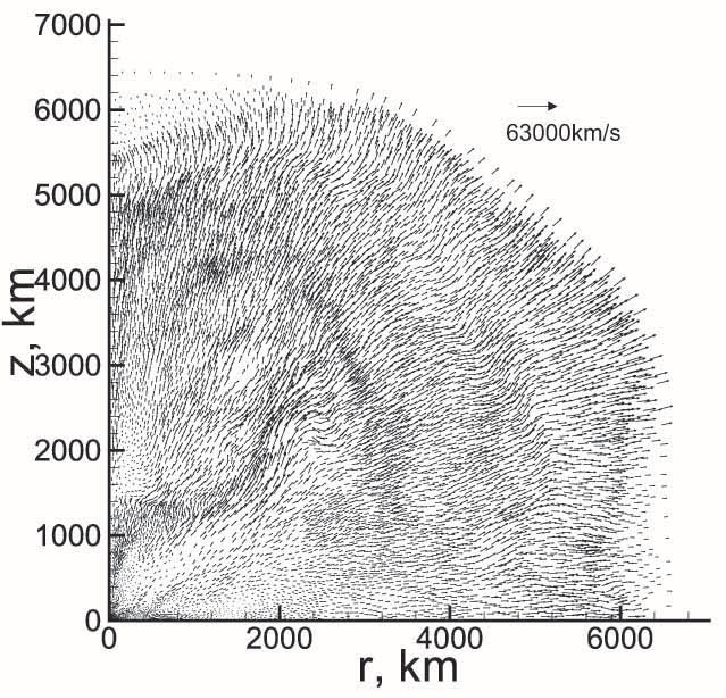}}
  \vspace{-0.3cm}
  \caption{Time evolution of the velocity field for time
   moments $t=0.075s,\> 0.1s,\> 0.25s$}.
  \label{vel1}
\end{figure}
\begin{figure}
   \centerline{\includegraphics{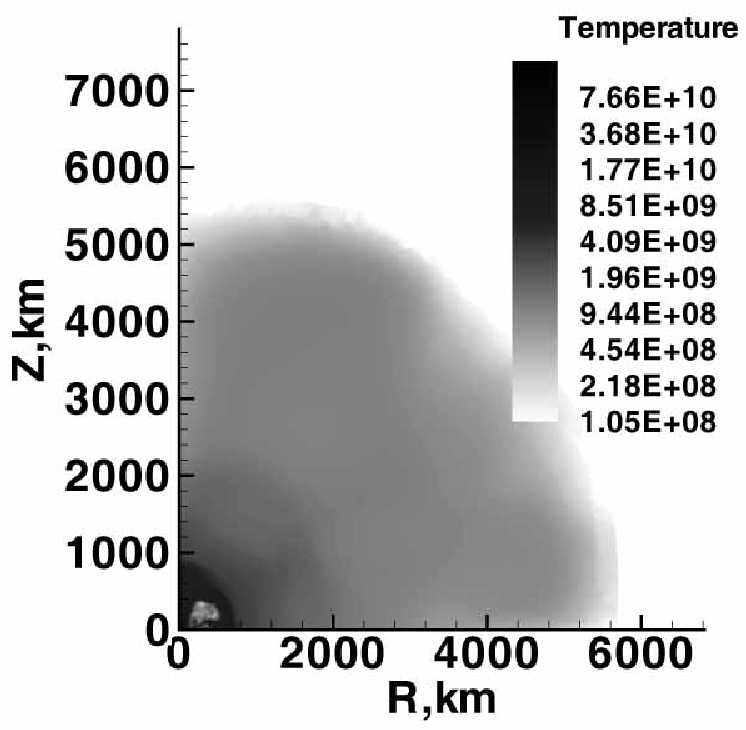}}
   \vspace{-0.2cm}
  \centerline{\includegraphics{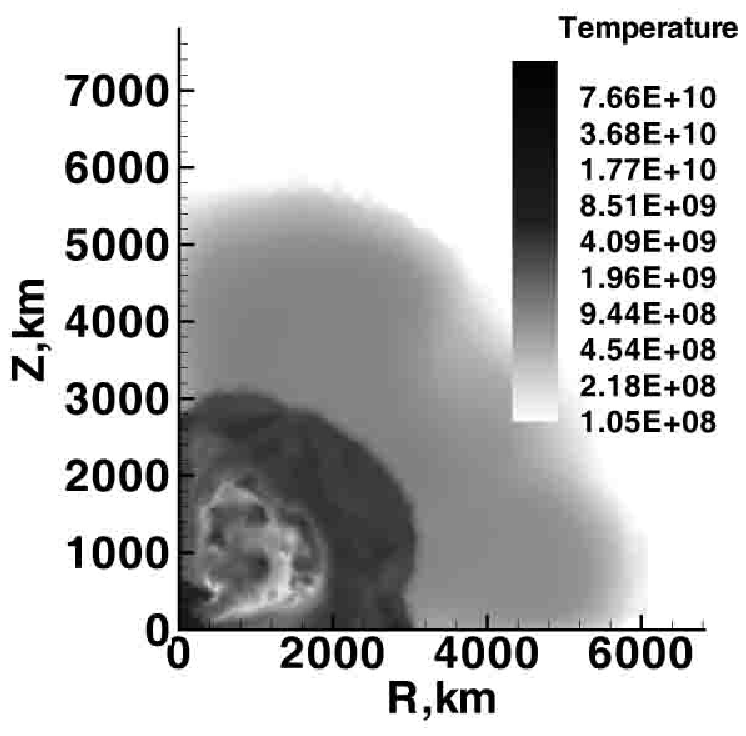}}
  \vspace{-0.2cm}
  \centerline{\includegraphics{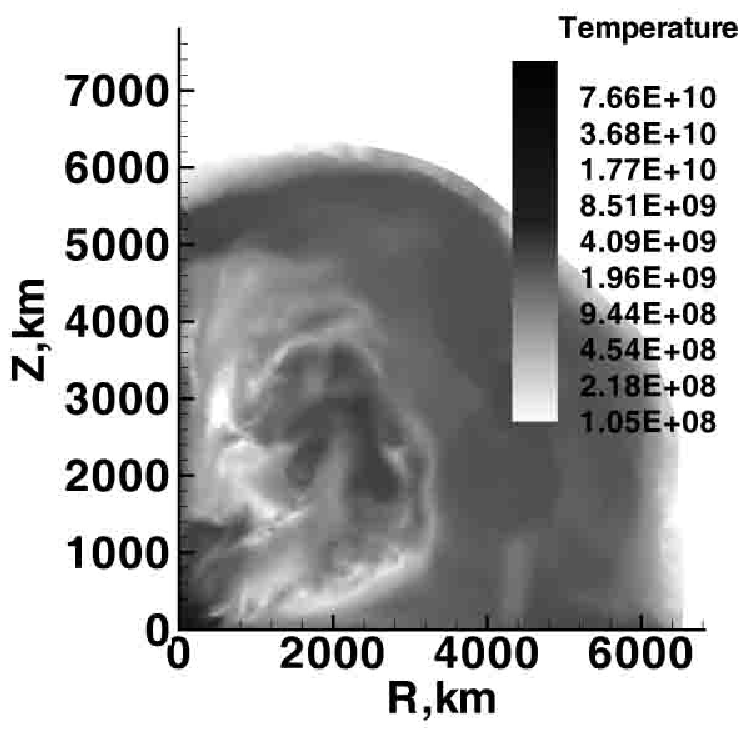}}
  \caption{Time evolution of the  temperature (in K)
   for time moments $t=0.075s,\> 0.1s,\> 0.25s$.}
  \label{temp1}
\end{figure}
\begin{figure}
  \centerline{\includegraphics{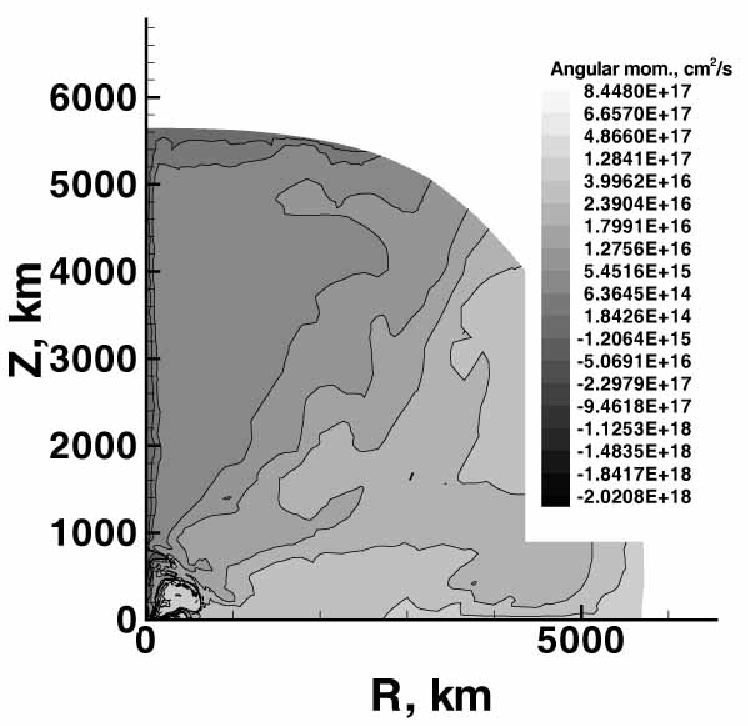}}
  \vspace{-0.2cm}
  \centerline{\includegraphics{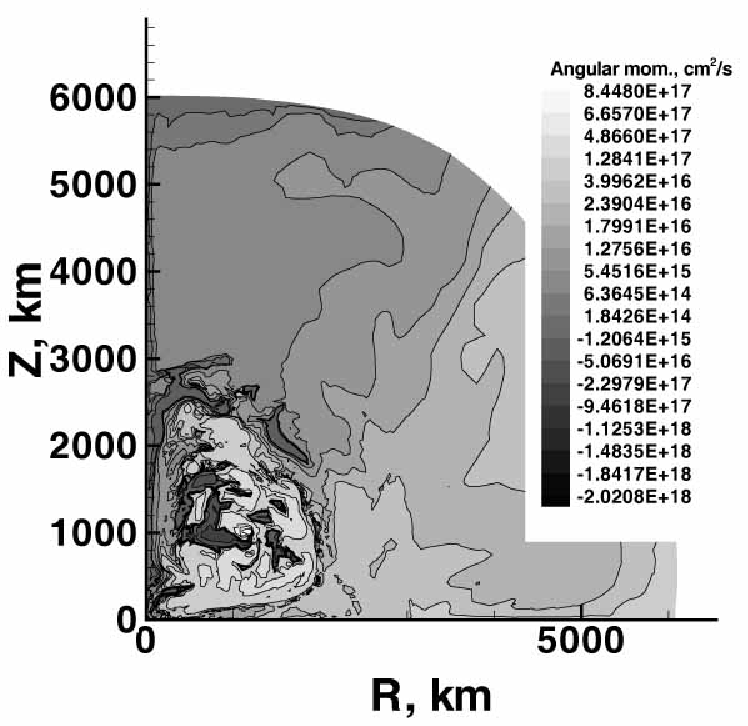}}
  \vspace{-0.2cm}
  \centerline{\includegraphics{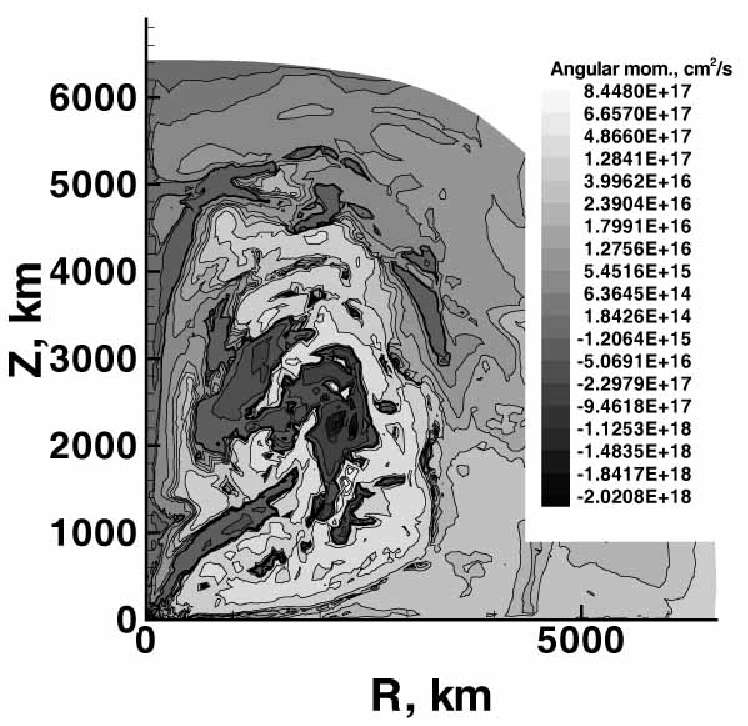}}
  \caption{Time evolution of the specific angular momentum $v_\varphi r$
  for time moments $t=0.075s,\> 0.1s,\> 0.25s$. }
  \label{angmom1}
\end{figure}

The time evolution of the velocity field and temperature are
represented in Figs.\ref{vel1} and \ref{temp1}. At the upper plot
of the Fig.\ref{vel1} a zoom of the central part of the velocity
field is represented. The shape of the supernova MHD shock is not
spherical. Near $z$ axis a protojet is forming. At the developed
stages of MR explosion the shape of the shock front becomes more
spherical, while the plot for the time evolution of the specific
angular momentum Fig.\ref{angmom1} shows that the angular momentum
is extracted mainly along the axis of rotation $z$.

A development of the supernova shock leads to formation of a
mildly collimated protojet. This protojet could be collimated
stronger during passing through the vast envelope of the star and
formation of the funnel flow.

As we have shown in the paper \citet{abkm2005} the MR mechanism
with the initial quadrupole-like magnetic field leads to the
explosion which develops preferably near the equatorial plane. The
shape of the MR
explosion qualitatively depends on the initial configuration of
the magnetic field. The magnetic field with a pure dipole or
quadrupole  types of symmetry in the star is a simplification. In
reality the structure of the magnetic field could be much more
complicated what means that the resulting shapes of the MR
supernova explosion could be rather different.

\begin{figure}
  \centerline{\includegraphics{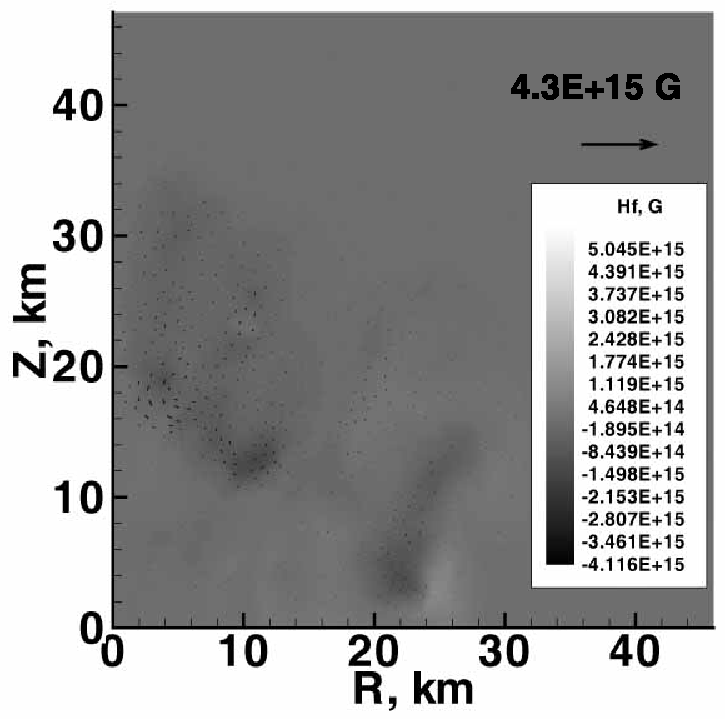}}
\vspace{-0.3cm}
  \centerline{\includegraphics{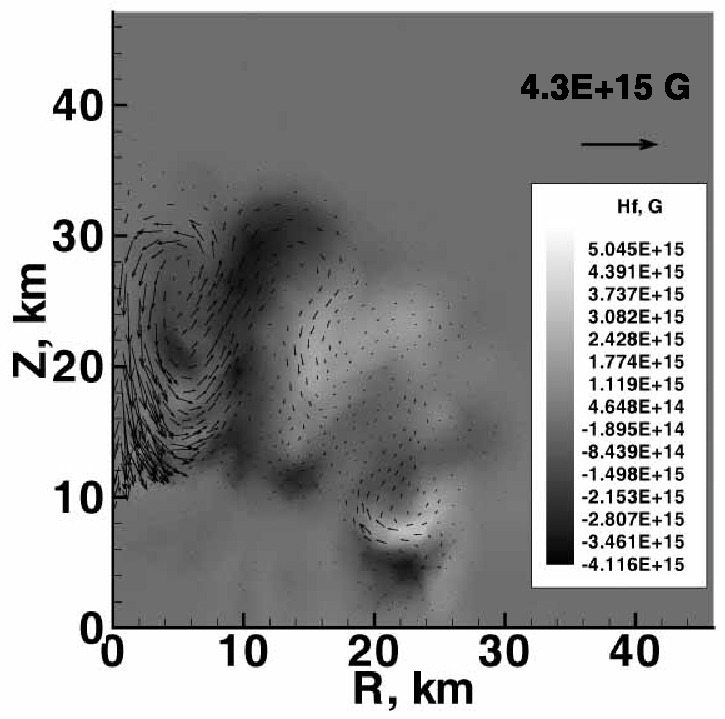}}
\vspace{-0.3cm}
  \centerline{\includegraphics{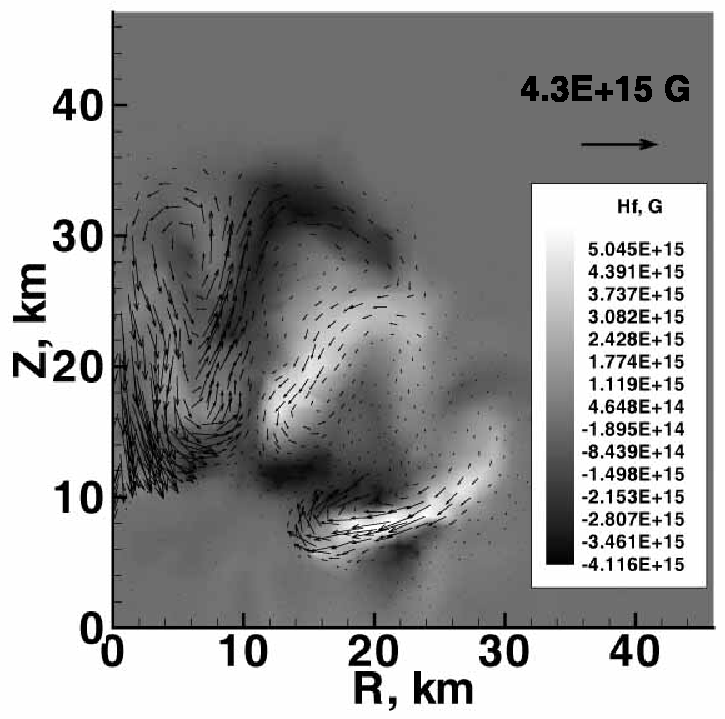}}
  \vspace{-0.2cm}
    \caption{The MRI development for time moments $t=0.0045s,\>0.018s,\>0.042s$.
    Gray scale is the toroidal field $H_\varphi$ levels. Arrows
    show a direction and strength
    of the poloidal magnetic field $H_r, \> H_z$.}
  \label{mmri1}
\end{figure}

\subsection{Appearance of the magnetorotational instability (MRI) in 2-D
picture}

It was shown in 1D simulations of MR supernova explosion
\citet{abkp} that the generated toroidal magnetic field is
growing linearly with the time. Our recent 2D MR supernova
simulations with the quadrupole-like initial magnetic field
(\citet{abkm2005}) have shown that,
after about 100 rotational periods of the central core,
its linear growth changes to the exponential one, and the poloidal
components also start to grow exponentially. The reason for that
exponential growth is an onset of the variant of MRI, which was
investigated by \citet{tayler}.
The qualitative explanation of MRI development for the MR supernova in
2D case was given by \citet{abkm2005}. Appearance of the MRI
significantly reduces the evolution time of the MR supernova
explosion.  The picture of the MRI development in the present simulations
is represented in Fig.\ref{mmri1}.

\begin{figure}
\centerline{\includegraphics{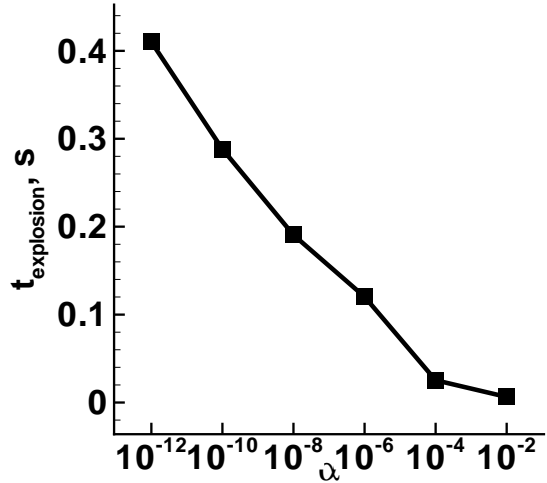}}
 \caption{The dependence of the time evolution of the MR explosion on
 $\alpha=\frac{E_{mag}}{E_{grav}}$ for the case of the
 initial dipole-like magnetic field.}
  \label{mri}
\end{figure}

Let us define $\alpha$
as  a
ratio between the magnetic $E_{mag0}$ and
 gravitational  $E_{grav0}$ energies at the moment of "turning on" of the
magnetic field.
  It was found in 1D simulations
that the time of the field amplification until
the MR explosion,  which is equal approximately to the time of the whole explosion
$t_{expl}$, depends on $\alpha$ as
$t_{expl}\sim\frac{1}{\sqrt{\alpha}}$.  The results of 2-D MR supernova simulations with
the initial quadrupole-like magnetic field  have been represented in the paper by
\citet{mbka2004} for a wide range of
variation of  $\alpha= 10^{-2}\div 10^{-12}$.
It was found, that for $\alpha < \sim 10^{-4}$ the growth of $t_{expl}$ with decreasing
$\alpha$ is becoming much slower, and may be well approximated by a logarithmic formula
$t_{expl} \sim |\log{\alpha}|$ for small $\alpha\ll 1$.
In the case of the dipole-like initial magnetic field we have made
simulations for the same range of variation of the $\alpha$
parameter, and have found a similar
dependence which is represented in
Fig.\ref{mri}.

The logarithmic functional dependence
at very small $\alpha$ has a simple qualitative explanation.
First, we should have in mind, that the development of MRI starts at
$t=t_{mri}$, approximately
at the same moment for {\it all} initial values of
the magnetic field. The MRI development begins in the part of the star,
where the ratio of the toroidal and poloidal magnetic fields reaches
a definite value $f \sim$ few tens, what happens after about
100 rotational periods of the inner core. In another words,
the time of the field growth until the beginning of MRI
$t_{mri}$ does not depend on $\alpha$, if $\alpha$ is small enough. At larger $\alpha$
the explosion starts before the beginning of MRI.

Second, we should take into account that MR explosion happens approximately at the same
ratio $F<1$ of the magnetic $E_{mag}$ and internal $E_{int}$ (gravitational $E_{grav}$)
energies of the star, also for all $\alpha$, $E_{magE}=F E_{grav}$.
The magnetic energy $E_{mag}$ is growing exponentially with time
during MRI development at $t>t_{mri}$,

\begin{equation}
\label{mri1}
\frac{E_{mag}}{f E_{mag0}} = \exp[\gamma_m (t-t_{mri})],
\end{equation}
 where $\gamma_m$ is an increment
of the MRI instability. Therefore, the
time between the the start of MRI,
until the moment of the explosion $t_{expl}$, is growing logarithmically
with decreasing $\alpha$

$$
t_{expl}-t_{mri} = \frac{1}{\gamma_m}\log\frac{E_{magE}}{f E_{mag0}}
=\frac{1}{\gamma_m}\log{\frac{F E_{grav}}{f E_{mag0}}}
$$
\begin{equation}
\label{mri2}
\approx \frac{1}{\gamma_m}\log{\frac{F E_{grav0}}{f E_{mag0}}}
=\frac{1}{\gamma_m}(\log{\frac{F}{f}}-\log \alpha).
\end{equation}
Here it was taken into account that a relative change of the
gravitational energy of the star during MagnetoRotational
Explosion (MRE) is rather small, and $E_{grav}(t) \approx
E_{grav0}$. At larger $\alpha$ the MRE happens before MRI
instability begins, $t_{expl}<t_{mri}$, with $t_{expl}\sim
1/\sqrt{\alpha}$, and for small $\alpha$, when
$t_{expl}>>t_{mri}$, the explosion time is growing
logarithmically. This qualitative explanation is valid if the
explosion happens during a linear stage of the MRI development. On
this stage the magnetic field is growing exponentially with the
increment roughly defined by the values of the configuration at
the beginning of instability. When the magnetic field energy
roughly approaches the rotational energy of the star, its growth
is stopped, resulting in nonlinear saturation of MRI. Our
numerical simulations have shown, that the explosion occurs before
the nonlinear saturation of MRI.

\subsection{Influence of numerical dissipation, and magnetic field
reconnection}\label{sect_mri}

We have done MR supernova explosion simulations with
the initial dipole-like magnetic field (Fig.\ref{inimag}) at a
different number of knots in our triangular grid, starting
from 1500 up to 18000 knots. We have found that the
moment of the onset of the MRI depends visibly on the
number of grid points. The less knots we have in the grid, the
longer time it takes for starting the MRI. We
have found that the results of the simulations (namely the time of the
development of the MR supernova explosion) do not change when the
total number of the knots $N_n$ exceeds 15000. The results of the
simulations described here are obtained with the grid containing
about 15000 knots. The dependence (converging) of the time of the
explosion $t_{expl}$ on the number of knots has a pure numerical origin,
and is connected with a numerical dissipation. Such dissipation has a
stabilizing influence on the MRI, and its onset is shifted to larger values
of $f$ - the ratio of the toroidal and poloidal magnetic fields. Therefore, the
total time of MRE is increasing with decreasing of the number of knots,
due to increasing of the time $t_{mri}$. Numerical dissipation influence also
on the field growth during MRI development, decreasing the increment $\gamma_m$,
giving the same effect.
As follows from our simulations the influence of the numerical viscosity
becomes unimportant at $N_n \ge 15000$, when the time of viscous (numerical)
dissipation starts to exceed all characteristic times of MR explosion.


In the simulations we supposed that the matter of the star has infinite
conductivity, although in reality the conductivity is finite. In the
case of a finite conductivity and chaotic structure of the magnetic
field the reconnection of the magnetic field could be
important, leading to a real physical dissipation.

The theory of the magnetic field reconnection
is not complete up to now.
To analyze the importance of the magnetic field
reconnection in MRE, let us
 estimate (at least roughly) a
characteristic time for this process.
The growth of the magnetic field is accompanied by
appearance of the chaotic small-scale structure of the magnetic field. If the
characteristic reconnection time in smaller scales
is comparable with characteristic time
$\sim 1/\gamma_m$
 of the magnetic field amplification due to MRI action,
 then reconnection could lead to a
 delay in the magnetic field growth, and hence
 could decrease the efficiency of MR explosion.

For the estimation of
the characteristic time for the magnetic field reconnection we use
the formula for the reconnection time of the quickest  - Petschek-type
reconnection (\cite{priest}, \cite{galeev}):

\begin{equation}\label{rec}
  \tau_{reconn}=\frac{4({\rm ln (Re_m)} +0.74)}{\pi
  v_Al^{-1}},
\end{equation}
where $v_A=H/(4\pi\rho)^{1/2}$  is the Alfv$\acute{\rm e}$n speed,
${\rm Re_m}=4\pi\sigma u l/c^2$  is the magnetic Reynolds number, $l$
 is the length, $u$ is the velocity, $\rho$ is the density,
$\sigma$ is the microscopic plasma conductivity,
$c$ is the light speed. In the region
where the MRI is developing we have the following parameters: $u=0.2\cdot
10^{10}$ cm/s, $l=10^5$cm, $\rho=7\cdot 10^{12}{\rm g/cm^3}$,
$H=0.17\cdot 10^{14}$G.

The plasma conductivity due to binary collisions $\sigma$ is defined as
(see e.g. \citet{croll})

\begin{equation}\label{cond}
    \sigma=\frac{3 m_e}{(16 \sqrt{\pi})Ze^2 {\rm ln}
    \Lambda}\left(\frac{2{\textrm \ae}T}{m_e}\right)^{3/2},
\end{equation}
where $m_e=9.1\cdot 10^{-28}$g is the electron mass, \ae$=1.38\cdot 10^{-16}\textrm{erg K}^{-1}$
is the Boltzmann constant, $e=4.8\cdot 10^{-10}$  is the electron charge, $Z=27$  is the ion charge,
${\rm ln}\Lambda=10$  is the Coulomb logarithm, $T=10^{11}$K  is the temperature.
 The above expression gives the conductivity of electrons in a non-relativistic, non-degenerate
plasma. Relativity may be taken approximately by substituting the light  speed
$c^3$ instead of the expression in brackets in (\ref{cond}), and increasing the
electron mass in the numerator by relativistic factor $\gamma \approx 20$.
Substituting all these data in formula (\ref{cond}) we find that
conductivity is $\sigma \approx 8\cdot 10^{20}\textrm{s}^{-1}$.
Note, the degeneracy of the electrons increase its conductivity, so this estimation may
be considered as a lower limit for $\sigma$.
The magnetic Reynolds number in our case is $\rm Re_m \approx 10^{15}$.
Substituting these data into the formula (\ref{rec}) we
find the absolute lower limit for the characteristic reconnection time as

$$
\tau_{reconn}\approx 5\,\, {\rm s}
$$
Our numerical simulations show that the characteristic time of the
development of MRI is much smaller ($\sim 10$ times)
than the characteristic reconnection time. We can conclude that
the reconnection processes will not suppress magnetic field
amplification and hence MR supernova explosion. The formula
(\ref{rec}) for the characteristic time of the reconnection used
here for the Petschek-type reconnection is the lower limit for the
reconnection time. We can not assert that this type of the
reconnection will be realized in the case of MR supernova
mechanism while in the case of the realization of any other
reconnection model the reconnection time will be even larger.

\section{Discussion}
The results of the 2D numerical simulations of the MR supernova
explosion with the initial dipole-like magnetic field have shown
that MR
supernova explosion is sensitive to the magnetic field
configuration. MR supernova explosion with the
initial quadrupole-like magnetic field develops mainly near the
equatorial plane. In the case of the
dipole-like magnetic  significant
part of the ejected matter obtains a velocity
along the rotational axis. The total energy
of the explosion in the MR mechanism does not depend
significantly
 on the topology of the initial magnetic field. The
explosion energy for quadrupole-like magnetic field is approximately
equal to $0.61\cdot 10^{51}$ erg \citet{abkm2005}, and for the dipole-like magnetic field
its value is about ${\mathrm E_{ejected}}\approx 0.5\cdot 10^{51}$ erg.
The amount of the ejected mass is approximately the same in both cases
$M_{ejected}\approx 0.14M_\odot$. Comparison of the explosion
times for the dipole-like and the quadrupole-like
fields of the same initial magnetic energy, shows that
in the quadrupole case the explosion is developing faster. The time of the
explosion for the  $\alpha=10^{-6}$ is about $\sim 0.12$ s in
the dipole case, and is about $\sim 0.06$ s in the case of
the quadrupole.

In both cases we have done 2D simulations in a wide range of the
initial magnetic field strength. The parameter $\alpha$ (the ratio
between the initial magnetic energy and the gravitational energy
of the star at the moment of "turning on" of the magnetic field)
was chosen as

\begin{equation}
\alpha=\frac{E_{mag0}}{|E_{gr}|}= 10^{-2}\div 10^{-12}.
\end{equation}
Comparison of the results of 1D simulations from \citet{abkp}, with
a dependence of the MR explosion time on $\alpha$ as
\begin{equation}
t_{expl}\sim\frac{1}{\sqrt \alpha} \label{mri1d},
\end{equation}
with corresponding 2D results (Fig.\ref{mri}),
shows a qualitative difference between them.
The reason is a development of the MRI in 2D-case, which
reduces drastically the MR explosion time $t_{expl}$.
For the values of $\alpha > \sim 10^{-4}$ the dependence (\ref{mri1d})
holds approximately, while for smaller $\alpha$ values the MR explosion time
depends on $\alpha$ in the following way:

\begin{equation}
t_{expl} \sim |\log{\alpha}|.
\end{equation}
The axial protojet forming in our simulations is not narrow, while
propagating through the envelope of the massive star this protojet
could be collimated, or suffer mass entrainment and deceleration.

The collapse does not lead to strong amplification of the toroidal
field, because its time in the central parts is comparable with
rotation time. Besides, the initial configuration is probably
stable against MHD instabilities and (in reality) contains both
parts, toroidal and poloidal, of the magnetic field. The relation
between these components, and details of the initial angular
velocity distribution (uniform in our case) may lead to different
situations, when MRI starts  to develop during the collapse, or
after many rotational periods of the new born neutron star . Here
we investigate the second case, as the initial step to the
problem. The time separation of the hydrodynamic collapse from MRI
and MRE helped us to understand the physical picture of MRI
development, and to obtain the estimations for the energy
production in MRE.

We plan to extend our calculations to a more realistic case when
magnetic field will be included at the very beginning of the
collapse. We expect to obtain more complicated picture of
magnetohydrodynamical processes, which should not influence
strongly on the energy production, depending mainly on the initial
rotational energy and angular velocity distribution.

In the paper by \cite{sawai} a similar problem with rather strong
initial magnetic field was simulated, and, contrary to our
results, the authors did not find the development of MRI. Their
simulations of the supernova explosion with weaker magnetic field
did not lead to the development of MRI either and hence they did
not get supernova explosion for a weak initial magnetic field. The
authors  (\citealt{sawai}) claim that MRI did not develop in their
simulations due to the insufficient spatial resolution. The
absence of the MRI in their calculations most probably is
connected with a rather large numerical viscosity of the numerical
scheme they used. The numerical viscosity of the simulations
depends on characteristic size of the mesh. The bigger is the size
of the cell the larger is the numerical viscosity. In the section
\ref{sect_mri} we have discussed the change of the MRI appearing
time moment in dependence of the spatial resolution of our grid.
In the case when the spatial resolution of the grid is very rough
we should  not get MRI in the case of application of our numerical
method also.

It is known from the observations that the shapes of core collapse
supernovae are different. From our simulations it follows that MR
supernova explosion arises after development of the MRI. The
development of the MRI is a stochastic process and hence the
resulting shape of the supernova can vary.
 We may conclude that MR supernova
explosion mechanism can lead to different shape of the supernova.
It is important to point out that MR mechanism of supernova
explosion leads always to asymmetrical outbursts.

The simulations of the MR supernova explosions for the initial
quadrupole-like magnetic field  described in the paper by
\cite{abkm2005} and MR supernova explosion described in current
paper are restricted by the symmetry to the equatorial plane.
While in reality this symmetry can be violated due to the MRI,
simultaneous presence of the  initial dipole and quadrupole -like
magnetic field (\cite{wang}) and initial toroidal magnetic field
(\cite{bkm1992}). The violation of the symmetry could lead to the
kick effect and formation of rapidly moving radio pulsars.

When rotational and magnetic axes do not coincide the whole
picture of the explosion process is three dimensional.
Nevertheless, the magnetic field twisting happens always around
the rotational axis, so we may expect the kick velocity of the
neutron star be strongly correlated with its spin direction, also
due to the anisotropy of the neutrino flux \citet{bk93}.
Simultaneously, because of the stochastic nature of MRI, the level
of the anisotropy should be strongly variable, leading to a large
spreading in the the neutron star velocities. This prediction of
MR explosion differs from the models with a powerful neutrino
convection, where arbitrary direction of the kick velocity is
expected (\citet{bur}). Recent analysis of observations of pulsars
(\citet{johnston}) shows, that rotation and velocity vectors of
pulsars are aligned, as it is predicted by the MR supernova
mechanism. The alignment of the vectors can be violated in the
case when supernova explodes in a binary system.

\section*{Acknowledgments}
The authors would like to thank RFBR  in the frame of the grant
No. 05-02-17697-a and the program "Nonstationary phenomena in
astronomy" for the partial support of this work. We would like to
thank Andrew Sadovskii for useful discussion.

\bsp

\label{lastpage}

\end{document}